\documentclass[finale]{Wiley}
\usepackage{graphicx}
\usepackage{float}
\usepackage{multicol}
\usepackage{color}
\usepackage{indentfirst}
\usepackage{amsmath} 
\usepackage{siunitx}
\usepackage{bm}
\usepackage{hyperref}
\usepackage{geometry}
\usepackage{ragged2e}
\justifying
\begin{document}
\pagestyle{fancy}
%\rhead{\includegraphics[width=2.5cm]{VCH-logo.png}}
\title{Controlling the Propagation of Flexural Elastic Waves With Ceramic Metatiles}
\maketitle

\author{
Brahim Lemkalli$^{1,*}$,
Ozgur T. Tugut$^{1}$,
Qingxiang Ji$^{1}$,
Richard Craster$^{2,3}$,
Sébastien Guenneau$^{3,4}$,
Muamer Kadic$^{1}$,
Claudio Bizzaglia$^{5}$,
Bogdan Ungureanu$^{3,4,6}$
}

\vspace{0.5cm}

\begin{affiliations}
$^1$Université Marie et Louis Pasteur, Institut FEMTO-ST, 25000 Besançon, France\\
$^2$Mathematics Department, Imperial College London, SW7~2AZ, London, UK\\
$^3$UMI 2004 Abraham de Moivre-CNRS, Imperial College London, SW7~2AZ, London, UK\\
$^4$The Blackett Laboratory, Physics Department, Imperial College London, SW7~2AZ, London, UK\\
$^5$IRIS, Via Guido Reni 2E, 42014 Castellarano, Italy\\
$^6$METAMAT Limited, 937 Chelsea C, Sloane Avenue, SW3~3EU, London, UK\\
$^{*}$lemkallibrahim@gmail.com
\end{affiliations}

\begin{abstract}
{In this work,} we examine the application of {phononic} metamaterials for {elastic} impact noise insulation in tiled {flooring, through the development of an innovative ceramic metatile that incorporates phononic crystals with optimized joint configurations.} First, we optimize the geometrical and material parameters of the proposed metatile, which is composed of small ceramic subtiles connected by silicon joints, in order to reduce longitudinal and flexural wave propagation on tiled floors, which are responsible for noise vibrations in tiled environments.  {A bandgap is achieved that effectively suppresses the transmission of impact noise through the periodic structural configuration. For flexural waves, the ceramic metatile exhibits a pronounced attenuation of wave transmission in the range of $500$–$1900$ Hz along the $[100]$ direction, and $500$–$1400$ Hz along the $[110]$ direction. For longitudinal waves, a broad bandgap is observed, spanning from $400$ Hz to $1950$ Hz in both the $[100]$ and $[110]$ directions. Additionally, the bandgaps shift toward lower frequencies with increasing width of the subtiles and silicon joints, or with a decrease in the Young’s modulus of the silicon.} In both experimental and numerical tests, it is demonstrated that the integration of silicon joints inside the ceramic metatile improves the acoustic insulation performance, as measured by the reduction of impact noise levels across a wide range of low frequencies. The findings highlight the potential of metamaterials in architectural acoustics, offering innovative solutions for elastic wave control in tiled environments.
\end{abstract}

%\begin{keyword}
%Impact elastic noise insulation\sep Tiled ceramic floors \sep Ceramic Metatiles\sep
%Phononic metamaterials\sep Low-frequency insulation\sep  broadband gaps
%\end{keyword}

\section{Introduction}
One of the most significant factors in building design is to provide a comfortable environment for its residents. Several elements influence a home's degree of comfort, including temperature, humidity, noise, air quality and hygiene. Neglecting such living conditions will have a negative impact on household comfort, e.g., noise generated by footsteps and noise pollution from surrounding buildings, temperature rise due to heat absorbed by buildings during the daytime, inadequate sanitation, and poor indoor air quality \cite{zhang2024acoustic,idrissi2025parametric, oquendo2022systematic}. Among these living conditions, noise control is a critical aspect of architectural design and urban planning, with far-reaching implications for physical and mental health \cite{pretzsch2021health, bartel2021possible}. Reducing noise in enclosed spaces is a long-standing objective in architectural acoustics, especially in contexts like tiled rooms, where vibration reflections and transmissions are enhanced by hard, reflective surfaces \cite{sharma2022review, wu2025reducing}. In general, there are numerous ways of controlling noise annoyance both within and outside buildings. Typical noise mitigation methods, such as the use of porous absorbers or heavy insulation, are often encountered with aesthetic, functional, or structural constraints. Recent advances in metamaterials \cite{kadic20193d, valipour2022metamaterials, jin20252024, barhoumi2022improved, dudek2025active, wang2024multistep, kadic2019static, wang2022nonlocal, lemkalli2025w, wang20223d, lemkalli2025space, ji20214d, lemkalli2023mapping, wang2023non}, including phononic crystals and acoustic metasurfaces, have opened new avenues for elastic waves control in buildings, enabling the manipulation of wave propagation using compact and tunable designs \cite{sharma2022review, sun2024sound}. Thus, the solution to improve the comfort of noise insulation in building floors should be based on phononic metamaterials in one part and institutive materials in the other part \cite{gibson2022low, oudich2023tailoring, muhammad2021phononic}. 

Phononic metamaterials, composed of periodic arrangements of scatterers and resonators with contrasting acoustic properties, are particularly promising due to their ability to create low-frequency band gaps that inhibit the propagation of specific wave modes \cite{jia2024maximizing, lu2009phononic, gross2023tetramode, wang2022nonlocal, luo2024bandgap,chen2020isotropic, zeng2022seismic}. These band gaps arise from mechanisms such as Bragg scattering and local resonances, which can be engineered to effectively control the propagation of longitudinal, transverse, torsional and flexural wave modes \cite{jin20252024, lemkalli2024longitudinal, lemkalli2024emergence, qu2024chiral}. Such capabilities make phononic metamaterials ideal candidates for integration into structural elements (e.g., ceramic tiles) to address noise control challenges in architectural applications \cite{maldovan2013sound, hussein2014dynamics, lemkalli2023lightweight, zhao2021acoustic, lemkalli2023bi, iglesias2021three, lemkalli2023innovative, craster2023mechanical}.

Numerous studies have explored the potential of acoustic metamaterials for vibration and impact noise control \cite{khodavirdi2024atomistic, fan2024multi, liu2024broadband, luo2024surface, jiang2023waveguides, chen2023multifunctional}. For example, Maldovan demonstrated the tunability of acoustic band gaps in phononic structures, emphasizing their role in mitigating low-frequency noise \cite{maldovan2013sound}. Similarly, Hussein et \textit{al}. reviewed wave attenuation mechanisms in phononic structures, underlining their relevance for multi-scale noise control designs \cite{hussein2014dynamics}.  {Ding et \textit{al}. investigated the effect of unit cell arrangement in a phononic array, introducing the concepts of isotacticity and syndiotacticity to achieve low-frequency and broadband band gaps \cite{ding2024isotacticity}. Zhou et \textit{al}. investigated an innovative method for controlling longitudinal wave band gaps, based on the application of an inhomogeneous and periodic thermal field \cite{zhou2021band}. Wang et \textit{al}. developed elastic dissipative metasurfaces for the broadband mitigation of Rayleigh waves \cite{wang2025dissipative}. In general, the works mentioned achieve band gaps through complex or tunable structures. However, there remains a strong demand for simpler and more adaptable phononic metamaterial that can provide broadband band gaps at low frequencies \cite{carrillo2023identifying}. } {Gao et \textit{al.} achieved two key objectives using phononic metamaterials with time-delayed feedback control. One maximizing the width of the bandgap at a given attenuation rate, and two achieving the highest possible attenuation rate within the targeted frequency range. } In architectural contexts, research has increasingly focused on embedding phononic metamaterials into building components. Zhao et \textit{al}. showed that periodic structures integrated into metamaterial panels can effectively absorb transmitted noise in the 400–800 Hz range \cite{zhao2021acoustic}, while Ghaffarivardavagh et \textit{al}. employed acoustic metamaterials to achieve low-frequency attenuation in compact spaces, specifically within the 450–500 Hz range \cite{ghaffarivardavagh2019ultra}. {More recently, Lu et \textit{al}. introduced an ultrathin nonlocal acoustic metaliner, designed using a mode-matching method, which enables global resonance modulation and effectively suppresses anti-resonance between resonance modes, thereby mitigating broadband fan noise. The metaliner operates efficiently in the frequency range of $1000–5000$ Hz \cite{lu2025ultra}. Xu et \textit{al}. investigated metamaterial-based absorbers capable of simultaneously absorbing sound and vibration within the peaks' frequency at $850$ Hz and $1210$ Hz \cite{xu2024metamaterial}. Zhu et \textit{al}. studied a multilayer structure designed for low-frequency energy dissipation. They demonstrated that the absorption bandwidth can be broadened by adding more units, for instance, a two-unit multilayer structure achieved absorption in the range of 370 to 520 Hz\cite{zhu2023multilayer}.}

\begin{figure*}[ht!]
    \centering
    \includegraphics[width=0.9\linewidth]{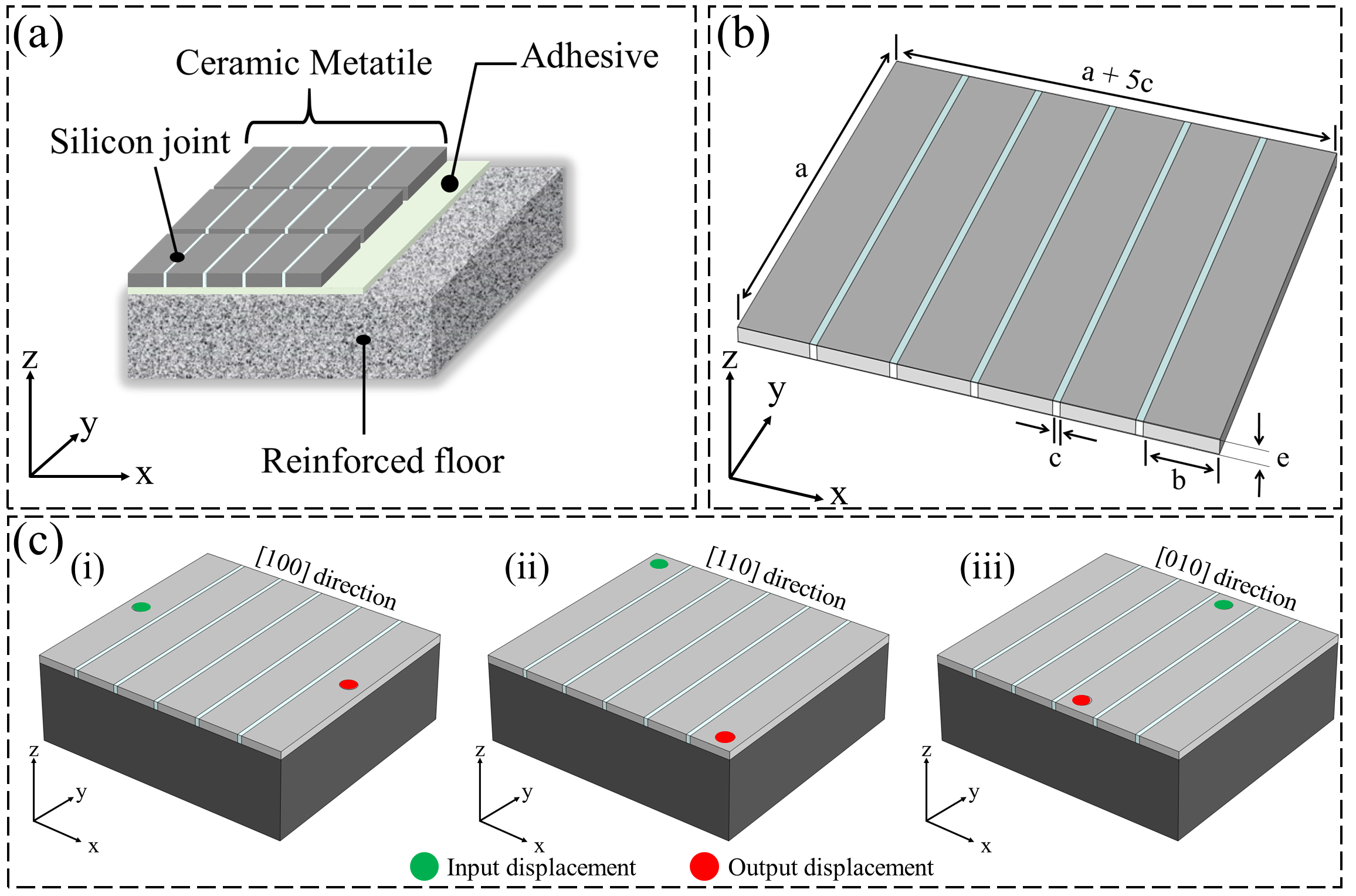}
    \caption{Schematic illustration of the proposed ceramic metatile. (a) Overview of the design strategy for impact noise insulation on ceramic metatiled floor. (b) Isometric view of the ceramic metatile with geometrical parameters. (c) Numerical model for transmission calculation, where the green dot represents the input displacement and the red dot represents the output displacement: (i) along the $[100]$ direction, (ii) along the $[110]$ direction, and (iii) along the $[010]$ direction.}
    \label{Figure 1}
\end{figure*}

Another key aspect of impact noise mitigation is wave absorption. Traditional absorptive materials rely on viscous and thermal dissipation, but recent studies have shown that phononic metamaterials can achieve efficient elastic wave absorption through localized resonances \cite{mei2012dark, sun2025inertial, yang2015broadband}. For instance, Mei et \textit{al}. demonstrated that hybrid structures combining phononic metamaterials with acoustic absorbers could significantly enhance performance across frequencies from 100 to 1000 Hz \cite{mei2012dark}. More recent developments have introduced broadband absorptive designs using sub-wavelength resonators, paving the way for compact and effective noise mitigation solutions \cite{yang2015broadband, huang2023towards, ke2024locally}.

Ceramic tile is a widely used construction material applied in various parts of buildings, including household roofs, walls, floors, and partitions \cite{berto2007ceramic, almeida2016environmental, fu2024recycling}. In modern architecture, floor tiles are often designed for durability, resistance to abrasion, water impermeability, ease of cleaning, and aesthetic appeal \cite{sudol2021makes}. However, in terms of acoustic performance,  ceramic tiles are not effective at reducing noise or vibrations \cite{lou2023enhanced, lou2023enhanced, amran2021sound}. Due to their high density and rigid surface, they are less effective at attenuating elastic vibration transmission compared to materials such as wood \cite{zhao2023scalable}, laminate\cite{gu2022laminated}, or rubber \cite{zanatta2021eco}.

To address the issue of noise insulation in tiled surfaces, several studies have proposed innovative solutions \cite{ma2025attenuation}. Kumar et \textit{al}. developed a metatile prototype based on a circular maze-like acoustic meta-structure, utilizing a space-coiling technique to enhance sound insulation in the low-to-mid frequency range $820–1600$ Hz. However, this design primarily targets airborne sound rather than impact noise generated by footsteps on floors \cite{kumar2023reconfigurable}. Choi et \textit{al}. investigated locally resonant metamaterials integrated into flooring systems of under-construction residential buildings, focusing on vibro-acoustic control of floor impact noise \cite{choi2022development}. Lopes et \textit{al}. proposed a methodology to maximize broadband impact noise insulation by designing slabs with non-uniform material thickness, optimized through numerical simulations \cite{lopes2024ribbed}. Additionally, Sukontasukkul et \textit{al}. explored the integration of viscoelastic polymer sheets into ceramic tiles to enhance noise insulation, particularly at low frequencies. While the performance at low frequencies was similar to conventional tiles, the tiles with viscoelastic layers exhibited superior performance at high frequencies above 1000 Hz \cite{sukontasukkul2023use}. 

Despite these advancements, achieving effective insulation against impact noise, transmitted via elastic wave propagation, remains a significant challenge in ceramic-tiled floors \cite{reinosa2022challenge, lourenco2025challenges}. Many existing solutions rely on alternative materials or complex designs that may compromise the durability and aesthetic qualities inherent to ceramic tiles. Moreover, some approaches are not easily compatible with conventional tile applications \cite{yang2021survey}. Previous studies have demonstrated that tile size and material properties play a critical role in tuning bandgap frequencies for noise vibrations insulation applications \cite{dal2021flexural, fang2018tunable,zhang2023nonlinear, zhao2021acoustic, zhou2022impact}. In our study, we aim to investigate elastic wave propagation in ceramic-tiled floors, with a particular focus on controlling flexural waves through the application of phononic metamaterials. Additionally, silicon joints are utilized as coupling elements to optimize wave transmission and reflection, aligning with findings by Fang et \textit{al}. \cite{fang2018tunable} on the effectiveness of elastomeric joints in acoustic metamaterials.
%add comp

Building on these principles, this research explores the use of phononic metamaterials and silicon joint design to reduce impact elastic waves in tiled rooms. Inspired by the concept of metatiles \cite{kumar2023reconfigurable}, a term combining metamaterials and construction tiles, we propose a new design that integrates ceramic plates and silicon. This approach preserves the key properties of ceramic tiles, such as durability, abrasion resistance, water impermeability, ease of cleaning, and aesthetic appeal, while also improving their elastic waves reduction capabilities. First, we analyze and optimize the geometrical parameters and material properties of metatiles, interconnected by silicon joints, to suppress longitudinal and flexural wave propagation. Next, we calculate the dynamic response using the finite element method. Finally, we present experimental validation and evaluate the acoustic performance of ceramic metatiles in impact elastic waves insulation, demonstrating their potential for practical applications in architectural acoustics for tiled floors. 
{The results reveal broadband band gaps that cover the frequency range typically generated by footsteps and indoor noise. To characterize our design, we have structured the paper as follows: We begin by describing the design of the ceramic metatile, followed by the optimization of its geometrical parameters. We then present numerical results for two types of elastic wave polarizations, flexural and longitudinal, as detailed in Section \ref{sec1}. Next, we describe the laboratory setup used to experimentally measure the transmission of flexural waves, along with the data processing performed using the Fast Fourier Transform (FFT) in Section \ref{sec1}. Finally, we examine a real-world scenario involving the generation of elastic impact noise. The corresponding FFT analysis is presented in Section \ref{sec3}, demonstrating the effectiveness of our ceramic metatile-based design.}

\section{Theoretical Dispersion and Transmission}\label{sec1}
{This section presents the design methodology of the ceramic metatiles and the optimization of the geometrical parameters of their constituent units. Additionally, it describes the numerical methods used to calculate transmission for both longitudinal and flexural waves.}
\subsection{Design of the ceramic metatile}
The proposed ceramic metatile is a periodic structure composed of ceramic sub-tiles and silicon joints. \autoref{Figure 1}(a) shows an isometric view of the main principle behind using ceramic metatiles to mitigate elastic wave propagation from impact noise or footstep noise. As depicted, the metatile floor consists of ceramic metatiles adhered to reinforced concrete using an adhesive. The periodic arrangement of ceramic sub-tiles and silicon joints resembles a monoatomic chain of resonators, with the silicon joints acting as springs and the ceramic sub-tiles serving as masses. This design enables elastic wave control through the periodic mass-spring system, using Bragg scattering and local resonance between neighboring sub-tiles. The geometric parameters of the metatile are illustrated in \autoref{Figure 1}(b), where $a$ represents the periodicity constant of a standard ceramic tile, $c$ is the width of the silicon joints, $b$ is the width of the subtiles, and $e$ is the thickness of the ceramic tiles taken as $e = 1 \, \text{cm}$ in this study. The standard ceramic tiles, provided by IRIS, were characterized through static and dynamic tests under small strains, yielding  following estimates for their elastic properties: Young's modulus $E_c = 55 \pm 5 \, \text{GPa}$, Poisson's ratio $\nu_c = 0.3$, and density $\rho_c = 3000 \pm 300 \, \text{kg/m}^3$. For the silicon material, we use commercially available silicon with the following mechanical properties: Young's modulus $E_s = 1 \text{MPa}$, Poisson's ratio $\nu_s = 0.4$, and density $\rho_s = 1000 \text{kg/m}^3$.

\subsection{Numerical models}
To characterize the noise impact reduction of our structure, we conduct a series of numerical simulations using the finite element method in the commercial software COMSOL Multiphysics. Our goal is to investigate the effect of adding silicon joints to the standard tiles on impact and footstep noise insulation. In doing so, we account for the linear elastic properties of the material by implementing a numerical model based on the weak formulation of the time-harmonic Navier equation, as follows:
\begin{equation}\label{Eq01}
    \frac{E}{2(1+\nu)}\biggl(\frac{1}{(1-2\nu)}\nabla (\nabla.\textbf{u})+\nabla^2 \textbf{u}\biggl)=-\omega^2\rho\textbf{u},
\end{equation}
where $\textbf{u}=(u, v, w)$ is the displacement vector, $\omega$ is the angular wave frequency and $\nabla$ is the gradient, $E$ is the Young’s modulus, $\nu$ is the Poisson’s ratio.

First, we compute their phononic dispersion relations, assuming that the ceramic metatiles are repeated periodically along the $x$-direction and finite in the $y$-axis. We apply the Floquet-Bloch boundary conditions along the $x$-direction as:
\begin{equation}\label{eq02}
    \textbf{u}(x+d)=\textbf{u}(x)e^{ikd},
\end{equation}
where ${\bf k}=(k, 0, 0)$ is the reduced Bloch wavevector and $\textbf{u}(x)$ the displacement vector at point $x$. Cells are assumed to be periodically repeated along the $x$-direction with periodicity $d=a+5c$.

The phononic dispersion relations are calculated in the first irreducible Brillouin zone $(\Gamma X)$ in the reciprocal space, where $\Gamma = (0, 0, 0)$ and $X = (\pi/d, 0, 0)$. For the meshing, we take into account a maximum element size of $\lambda_{\text{min}}/10$ and tetrahedral elements. {To delineate the flexural waves, we computed the dispersion curves, taking into account the polarization weighting using the following expression:}
\begin{equation}\label{eq003}
    p_{\alpha}=\frac{\iiint_V |w|\; dxdydz}{\iiint_V \sqrt{|u|^2+ |v|^2+ |w|^2}\; dxdydz},
\end{equation}
{where $V$ is the volume of the elementary cell $|u|=\sqrt{u u^*}$ is the modulus of displacement vector, with $u^*$ the complex conjugate of $u$.}

Following our examination of the phononic dispersion in ceramic metatiles (infinite), we proceed to investigate their transmission characteristics through simulations involving a single metatile (finite). We compare the results for three different directions for both the standard tile and the metatile, as depicted in \autoref{Figure 1}(c). The transmission of the elastic waves is calculated using the following equation:
\begin{equation}\label{Eq02}
{\rm Transmision} \;{(\rm dB)}=20\log_{10}\left(\frac{|u_i|_{\rm output}}{|u_i|_{\rm input}}\right),
\end{equation}
where $|u_i|_{\text{input}}$ and $|u_i|_{\text{output}}$ denote the magnitude of the displacement vector component (i.e., $\sqrt{uu^*}$, $\sqrt{vv^*}$, or $\sqrt{ww^*}$) at the input and output of the ceramic metatile, respectively. The component $u_i \in \{u, v, w\}$ is selected based on the direction of the input wave and depends on the nature of the wave polarization: longitudinal or flexural.

\subsection{Numerical results}
The main focus of this study is the investigation of bandgaps, which indicates frequency ranges where the ceramic metatile effectively attenuates footstep and impact noise vibrations. {To get accurate and trustworthy results, it is important to create a clear method for improving the size and material features of the proposed ceramic subtile and its silicon joints. To achieve this, we conduct an optimization study aimed at obtaining broadband band gaps and determining the optimal geometrical and material parameters. The study is performed using COMSOL Multiphysics, with manual optimization carried out through a parametric sweep approach; one parameter is fixed while the other is varied. We begin by varying the width of the subtiles while keeping the other parameters constant to identify the optimal width. Once the optimal width is determined, it is fixed, and we proceed to vary the thickness of the silicon joints. Finally, we study the influence of the material properties of the joints by performing a parametric sweep on their mechanical characteristics by analyzing the bandgaps extracted from the phononic dispersion in the first irreducible Brillouin zone, specifically along the $ \Gamma X $ direction, for a unit cell consisting of two subtiles of width $ b $ and two silicon joints of width $ c $.}

\begin{figure}[h!]
    \centering
    \includegraphics[width=1\linewidth]{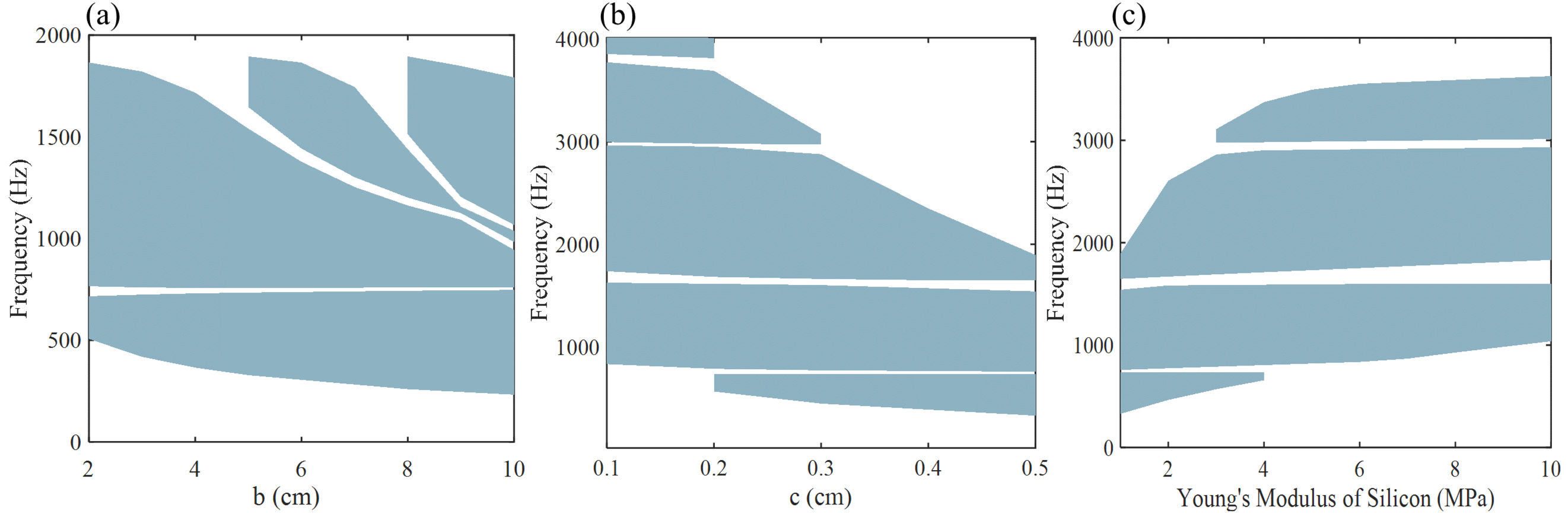}
    \caption{Directional bandgap presence as a function of geometric and material parameters extracted from the phononic dispersion in the first irreducible Brillouin zone along the x-direction. (a) Bandgap frequencies as a function of the ceramic plate width, with fixed $E = 1 \rm MPa$ and $c = 0.5 \rm cm$. (b) Bandgap frequencies as a function of the silicon width, with fixed $E = 1 \rm MPa$ and $b = 5 \rm cm$. (c) Bandgap frequencies as a function of the Young’s modulus of silicon, with fixed $c = 0.5 \rm cm$ and$ b = 5 \rm cm$.}
    \label{Figure 3}
\end{figure}

The bandgaps variations are examined as functions of the geometrical parameters of the ceramic metatile: the width of the subtiles $ b $ and the width of the silicon joints $ c $. Additionally, we investigate the influence of the Young's mosdulus of silicon on these bandgaps. The eigenvalue calculations are presented in \autoref{Figure 3} for an infinite structure along the $[100]$ and $[010]$ crystallographic directions.

Firstly, we fixed the width of the silicon joints at $ c = 0.5 $ cm and the Young's modulus of silicon at $ 1 $ MPa. Then, we varied the parameter $ b $ from $2$ cm to $10$ cm to analyze its effect on the bandgaps. \autoref{Figure 3}(a) presents the calculated bandgaps in function of $b$, which are colored in gray, along with their corresponding frequencies. It is observed that for $ b = 2 $ cm, the bandgaps start from $506$ Hz and extend up to approximately $1865$ Hz. As the width of the sub-tiles increases, the bandgaps shift toward lower frequencies, reaching a minimum of $232$ Hz. This behavior is consistent with the classical mass-spring system, where larger masses result in lower resonance frequencies due to the inverse relationship between frequency and mass. In such systems, both the mass and the stiffness of the spring play a critical role in determining the resonance frequency. Additionally, the increase in the sub-tile width $b$ influences Bragg scattering, as it alters the periodicity of the structure. In this case, the wavelength associated with Bragg scattering is approximately $\lambda=2\times b$; thus, as $b$ increases, the wavelength increases, leading to a corresponding decrease in frequency. This explains the observed downward shift in the bandgap frequencies with increasing $b$. This indicates that when the subtiles are larger, bandgaps appear in the lower frequency range. However, as their size continues to increase, the diffraction limit shifts further to lower frequencies, ultimately disrupting the bandgap around $1500$ Hz.

Secondly, for the variation of the silicon joint width $ c $, we fixed the width of the subtiles at $ b = 5 $ cm and the Young's modulus of silicon at $1$ MPa, as depicted in \autoref{Figure 3}(b). Then, we varied $ c $ from $0.1$ cm to $0.5$ cm to limit the size of silicon joints, ensuring that the ceramic metatile retains its resemblance to a standard one, with only small additional silicon joints. It is observed that for $ c = 0.1 $ cm, the bandgap starts at $831$ Hz. As the width of the silicon joints increases, this bandgap shifts toward lower frequencies, reaching $328$ Hz for $ c = 0.5 $ cm. This parametric study provides insights into how the silicon joints influence the frequency range of the bandgaps and how they can be controlled by introducing silicon joints into the ceramic metatile. However, the dimensions of the silicon joints also depend on the target frequency range for the intended application.

In the final parametric study, we fixed the dimensions of the ceramic metatile with a subtile width of $ 5 $ cm and a silicon joint width of $ 0.5 $ cm, as depicted in \autoref{Figure 3}(c). We varied the Young's modulus of silicon to better understand how silicon mechanical properties affect noise cancellation. By keeping the density and Poisson's ratio of the silicon constant, we varied the Young's modulus from $1$ MPa to $10$ MPa. It is observed that as the Young's modulus decreases, the bandgaps shift to lower frequencies. For example, for $ E = 1 $ MPa, the bandgaps start at $328$ Hz. As the silicon becomes stiffer, the bandgaps shift to higher frequencies. For $ E = 10 $ MPa, the bandgaps start at $1037$ Hz.

\begin{figure}[ht!]
    \centering
    \includegraphics[width=0.6\linewidth]{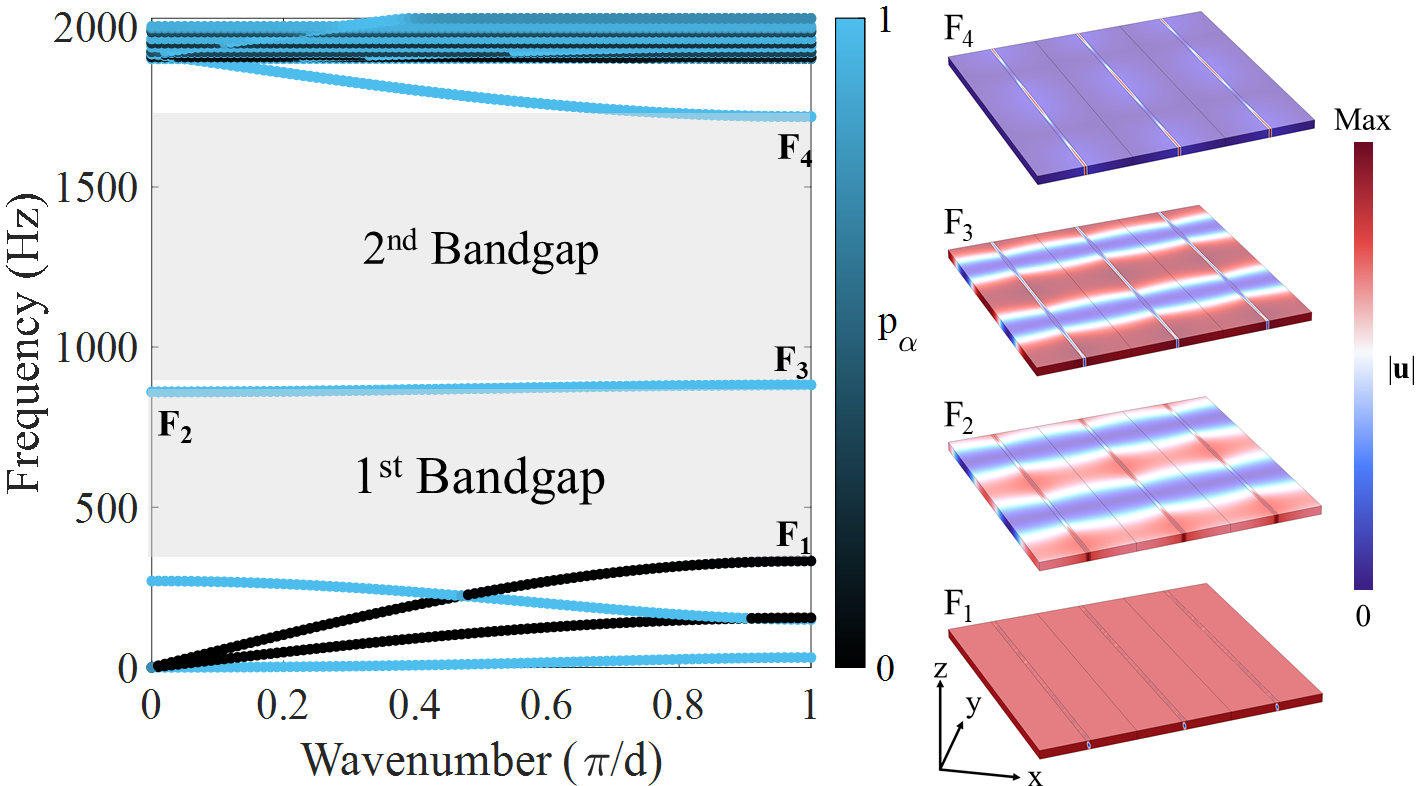}
    \caption{Phononic dispersion of the optimized ceramic metatile in the first irreducible Brillouin zone ($\Gamma X$). The geometric and material parameters used are $E = 1 MPa$, $c = 0.5 cm$, and $b = 5 cm$.}
    \label{figure 5}
\end{figure}

This analysis of the material and geometric parameters provides us with a better understanding of the bandgaps and their appearance in the ceramic metatile. After this detailed analysis, we chose $ c = 0.5 $ cm, $ b = 5 $ cm, and a silicon Young's modulus of $1$ MPa, which we consider for the rest of the paper. Using these values, we plot the dispersion curves in the first irreducible Brillouin zone along the $[100]$ crystallographic direction. The results of this study are presented in \autoref{figure 5}. {This figure shows the bandgaps, represented by the gray-colored region, and the propagating modes for the flexural waves, which are plotted in blue. Additionally, we display the eigenmodes of the ceramic metatile around the bandgaps, shown as screenshots of the total modulus displacement distribution for a ceramic metatile composed of $3$ unit cells. One can observe that the formation of the bandgaps affects all wave modes, meaning that within the bandgaps, no mode, whether flexural or longitudinal, can propagate. The first bandgap, starting from $F_1=332$ Hz  to $F_2=860$ Hz, is due to Bragg scattering, which is related to the condition $\lambda = 2\times b$, as depicted in the eigenmodes of the first-order mode vibration. Similarly, the second bandgap, from $F_3=882$ Hz to $F_4=1719$ Hz, corresponds to the eigenmode of the second-order vibration.} In general, the bandgaps are formed by both Bragg scattering and the resonance of the mass-spring system created by the silicon joints and subtiles, which causes energy localization at the resonance frequencies of the system.

\begin{figure}[h!]
    \centering
    \includegraphics[width=1\linewidth]{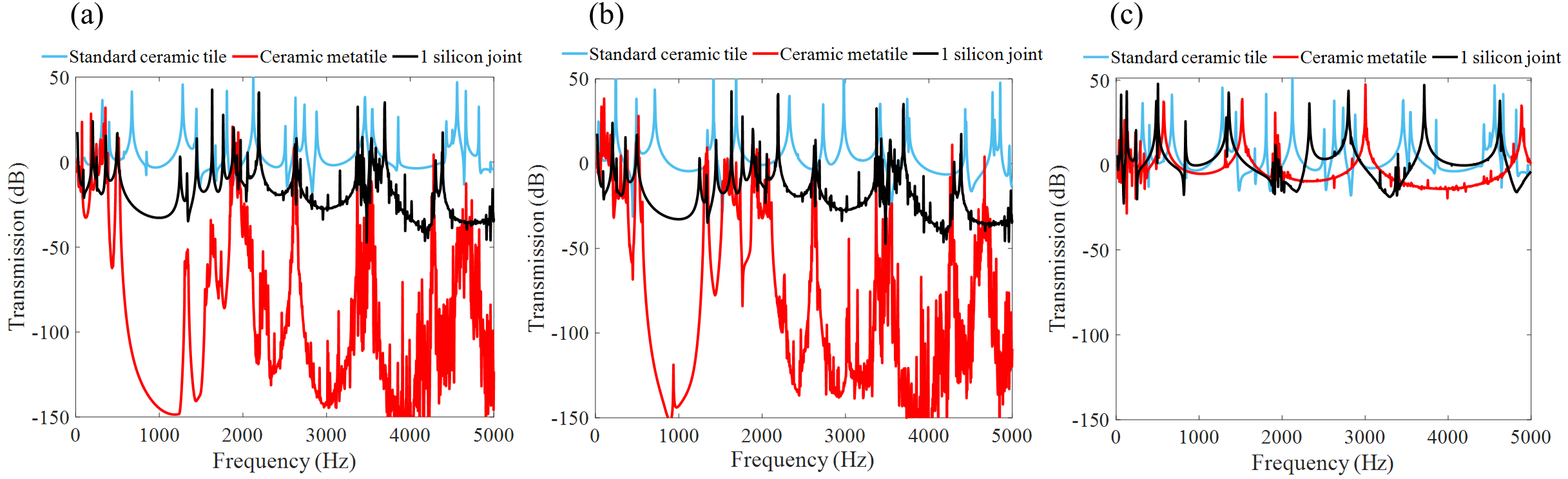}
    \caption{Numerical transmission curves for {flexural} waves. (a) Normal to the silicon joints along the $[100]$ direction. (b) Diagonal to the silicon joints along the $[110]$ direction. (c) Parallel to the silicon joints along the $[010]$ direction.}
    \label{figure 6}
\end{figure}

After calculating the phononic dispersion of the ceramic metatile, modeled as an infinite structure, we now proceed to characterize its dynamic behavior by analyzing the transmission of elastic waves through a single ceramic tile. This tile consists of $5$ silicon joints and $6$ ceramic subtiles, as illustrated in \autoref{Figure 1}(c). The transmission spectrum is calculated along the three principal directions: {$[100]$, $[110]$, and $[010]$,} to assess the ceramic metatile's response to wave propagation in different orientations.

\begin{figure}[h!]
    \centering
    \includegraphics[width=1\linewidth]{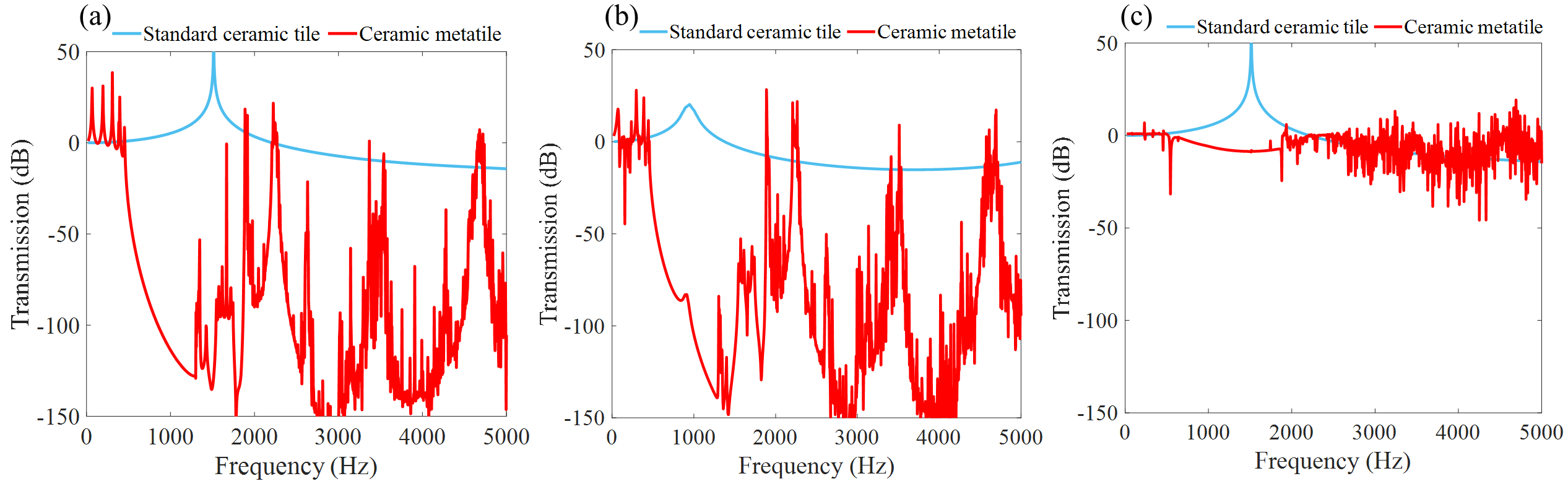}
    \caption{Numerical transmission curves for longitudinal waves. (a) Normal to the silicon joints along the $[100]$ direction. (b) Diagonal to the silicon joints along the $[110]$ direction. (c) Parallel to the silicon joints along the $[010]$ direction.}
    \label{figure 41}
\end{figure}

We begin by applying a {flexural} wave and calculating its transmission, as illustrated in \autoref{figure 6}. \autoref{figure 6}(a) presents the computed {flexural} transmission along the $[100]$ direction, where the red curve corresponds to the ceramic metatile, the blue curve represents the standard ceramic tile and the black curve illustrates the transmission through one silicon joint. In this direction, the transmissions of the standard ceramic tile and the $1$ silicon joint remain high across the entire frequency range studied ($20-5000$ Hz). In contrast, the ceramic metatile exhibits a significant reduction in transmission within the $500-1900$ Hz range, whereas in the lower range ($20-500$ Hz), the transmission remains nearly total. Beyond $1900$ Hz, the bandgaps are affected by diffraction limits, yet they retain their general shape and continue to attenuate vibrations effectively.

In the $[110]$ direction (\autoref{figure 6}(b)), the bandgap extends from $500$ Hz to $1400$ Hz, after which the transmission increases until $2000$ Hz. Beyond this point, additional bandgaps appear, influenced by the diffraction limit. Meanwhile, the standard ceramic tile maintains nearly maximal transmission across the entire frequency spectrum. When analyzing transmission in the $[010]$ direction (\autoref{figure 6}(c)), where wave propagation is parallel to the silicon joints, we observe that the ceramic metatile's transmission closely resembles that of the standard ceramic tile. This indicates that the presence of silicon joints in the propagation direction of {flexural} waves attenuates vibrations similarly to the inclined $45°$ joints. These findings suggest that our design effectively reduces {flexural} waves propagation in directions where the silicon joints are not aligned parallel to the wave propagation.

To further validate the dispersion curves in the first irreducible Brillouin zone for all existing modes, we now compute the transmission for a longitudinally polarized wave. Using the same model, we apply a longitudinal wave as the input displacement and measure the output displacement. The results of this study are illustrated in \autoref{figure 41}, where the transmission of the longitudinal wave is analyzed in three different directions. \autoref{figure 41}(a) shows the transmission along the $[100]$ direction. The standard ceramic tile exhibits excellent transmission of longitudinal vibrations, whereas the ceramic metatile displays significantly reduced transmission between $400$ Hz and $1950$ Hz. Beyond this range, bandgaps emerge, coupled with diffraction limits. Furthermore, we analyze the behavior of the inclined joints in relation to longitudinal wave propagation along the $[110]$ direction. Our design demonstrates a broad bandgap extending from $400$ Hz to $1950$ Hz, followed by the presence of bandgaps at higher frequencies, again influenced by diffraction effects, ad shown in \autoref{figure 41}(b). Finally, in the $[010]$ direction, the transmission behavior closely resembles that observed for {flexural} waves. In this case, both the standard ceramic tile and the ceramic metatile exhibit high transmission for both longitudinal and {flexural} waves, as depicted in \autoref{figure 41}(c).

We numerically demonstrate that our design, which incorporates silicon joints into ceramic metatiles, exhibits broadband bandgaps at low frequencies. Moreover, these bandgaps can be controlled by adjusting the geometrical and material parameters. The dispersion calculations presented in this section confirm that our chosen parameters create four distinct bandgaps in the low-frequency range of $20-2000$ Hz for an infinitely periodic structure. To further validate the presence of these bandgaps in a finite structure, we calculated the transmission of both {flexural} and longitudinal waves through a single ceramic metatile. Additionally, we numerically compared its response to that of a standard ceramic tile. In the next section, we present the experimental validation of these bandgaps.

\section{Experimental validation}\label{sec2}
{This section presents the experimental setup for measuring the transmission of flexural waves through the ceramic metatiles, along with the results obtained using the Fast Fourier Transform (FFT).}
\subsection{Laboratory transmission measurements}
\begin{figure}[ht!]
    \centering
    \includegraphics[width=0.7\linewidth]{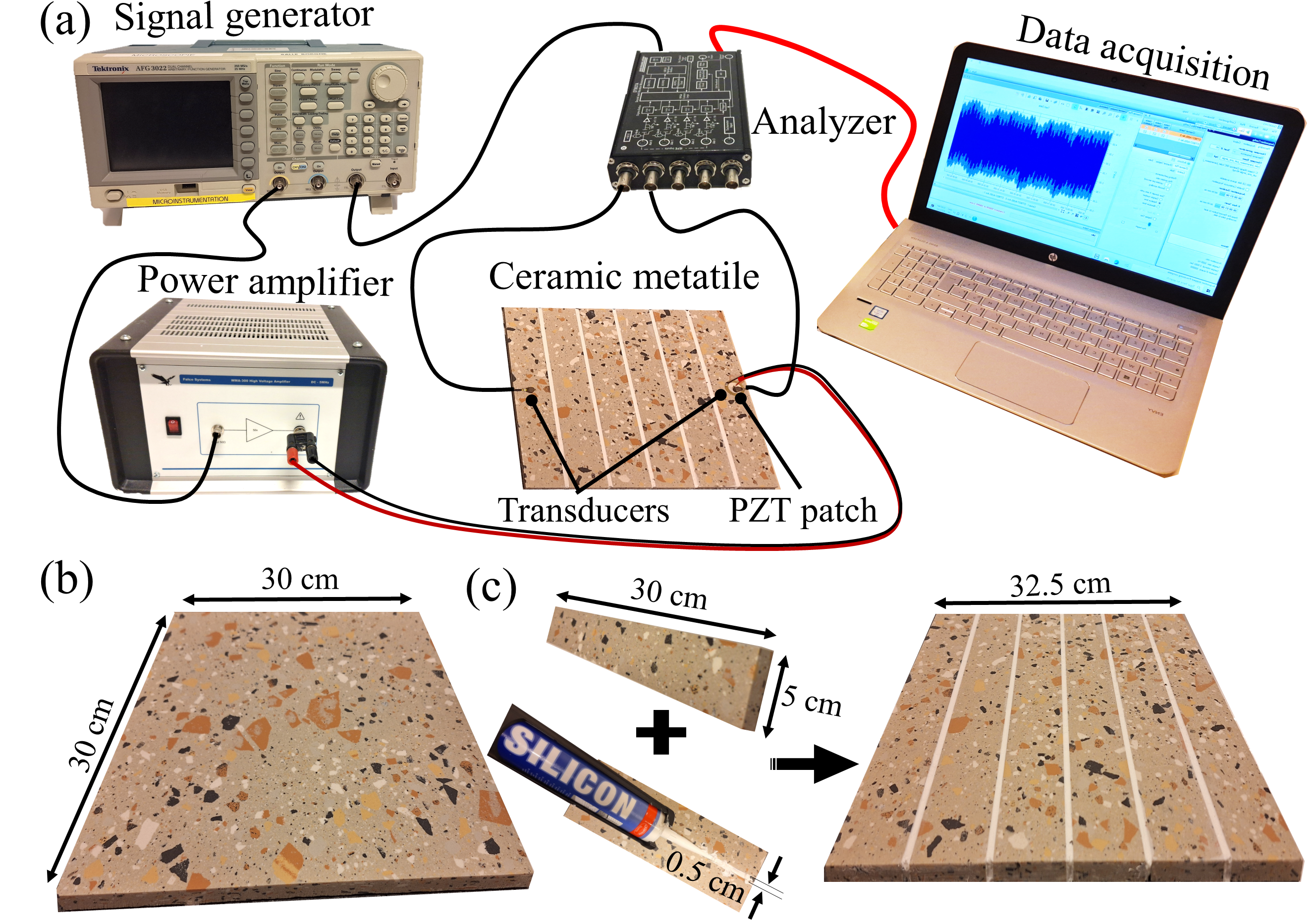}
    \caption{Illustrations of lab transmission measurements and sample fabrication. (a) A signal generator is used to shape the incident field before getting amplified and sent to the piezo patches. The signal is then traveling as an elastic wave and thus collected (read) by an accelerometer. Same accelerometer is glued on the piezo patch to read the truly generated elastic oscillation. The data are collected via a signal analyser and stored on the computer (b) The standard ceramic tile used in our experiments. (c) The ceramic metatile fabrication procedure, which involves silicon joint structures.}
    \label{figure 2}
\end{figure}
To validate numerical results, a ceramic metatile was fabricated by assembling ceramic plates with silicon joints, as shown in \autoref{figure 2}(c). The fabricated metatile has geometric parameters of width of $32.5 cm$ and $30 cm$, with a height of $1 cm$, which is composed of small ceramic plates of $5 cm$ as width and separated by silicon joint with a width of $0.5 cm$.

\autoref{figure 2}(a) illustrates the experimental setup used to measure the transmission spectrum of the ceramic metatile. A signal generator was employed to produce a sweep signal, which was fed into a power amplifier to enhance the signal amplitude. The amplified signal was then transmitted to a PZT patch, which generated the input displacement. A transducer measured this displacement, while another transducer at the output recorded the transmitted signal. The output transducer was connected to an analyzer, which interfaced with a laptop for data acquisition. This setup enabled the measurement and recording of both transmitted and incident displacements, allowing for the calculation of transmission curves.
\begin{figure}[ht!]
    \centering
    \includegraphics[width=1\linewidth]{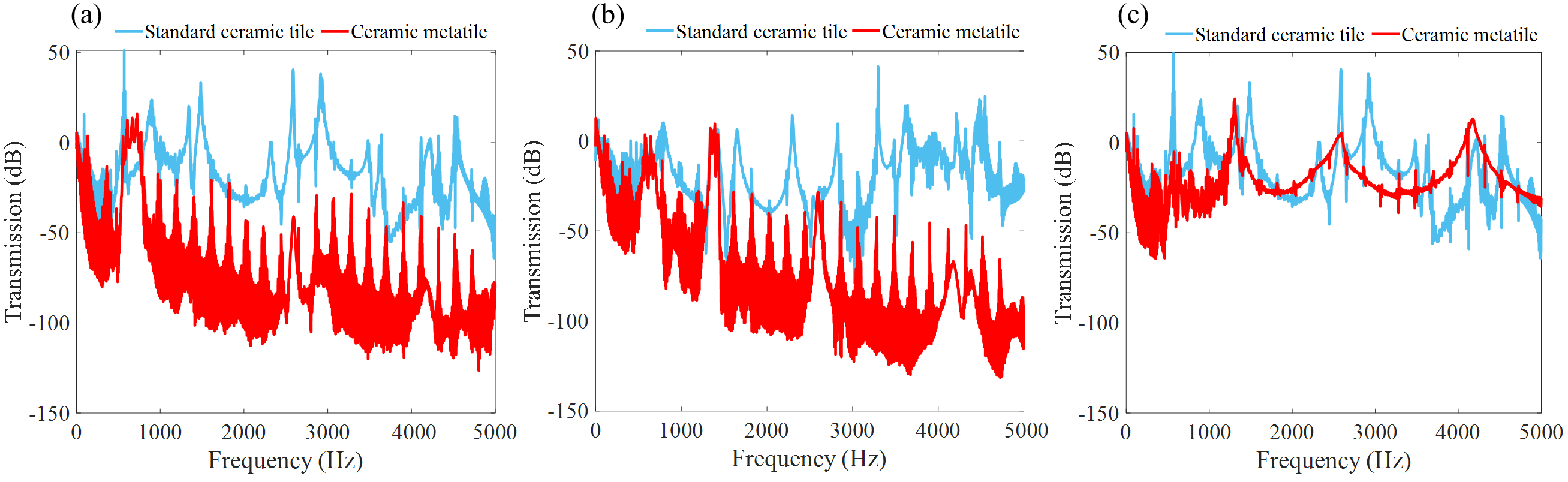}
    \caption{Experimental transmission curves for {flexural} waves. (a) Normal to the silicon joints along the $[100]$ direction. (b) Diagonal to the silicon joints along the $[110]$ direction. (c) Parallel to the silicon joints along the $[010]$ direction.}
    \label{figure 4}
\end{figure}
\subsection{Experimental results}\label{sec3}
For the experimental transmission measurements, we analyze the propagation of {flexural} waves along three directions: $[100]$, $[010]$, and $[110]$. Then, we compare the results obtained from a standard ceramic tile with those of our designed ceramic metatile. \autoref{figure 4} presents the experimental results, where the blue curves correspond to the standard ceramic tile, and the red curves represent the ceramic meta-tile's response in all directions. In the $[100]$ direction (\autoref{figure 4}(a)), the transmission remains high between $20$ Hz and $700$ Hz. Beyond this range, the transmission of the metatile decreases by approximately $50$ dB compared to the standard tile from $700$ Hz to $2500$ Hz. For the $[110]$ direction (\autoref{figure 4}(b)), a bandgap appears between $700$ Hz and $1400$ Hz. The transmission then increases until $1490$ Hz, after which the metatile exhibits a transmission reduction of around $50$ dB relative to the standard ceramic tile. finally, for the $[010]$ direction (\autoref{figure 4}(c)), both tiles exhibit high transmission of {flexural} waves. {Furthermore, the experimental results show good agreement with the numerical predictions in terms of both frequency ranges and attenuation amplitude. This is demonstrated by the calculated absolute error between the two results, given by the equation: $Error=\frac{T_{exp}-T_{num}}{T_{exp}}$ in which $T_{exp}$  the transmission at normal incidence of the flexural wave, and $T_{num}$ is its counterpart computed by FEM (see Appendix \ref{apA}.} {In addition, we perform numerical simulations to investigate the effect of damping by introducing an imaginary component to the Young’s modulus of the silicon joint, incorporating the loss factor $\eta$ as follows: $E_{\text{silicon}} = E(1 + i\eta)$. The transmission along the [100] direction is computed for three different damping scenarios in order to compare the numerical results with the experimental data (see Appendix \ref{apB}). We observe that increasing the damping in the silicon joint leads to broader band gaps and higher attenuation of the flexural waves.}

{To summarize, both the error analysis between experimental and numerical results, and the study of the silicon joint’s loss factor, confirm that the observed band gaps are associated with Bragg scattering and local resonances, and that the experimental results are in good agreement with the numerical predictions.
}

\section{Discussion}

\begin{figure*}[h!]
    \centering
    \includegraphics[width=0.9\linewidth]{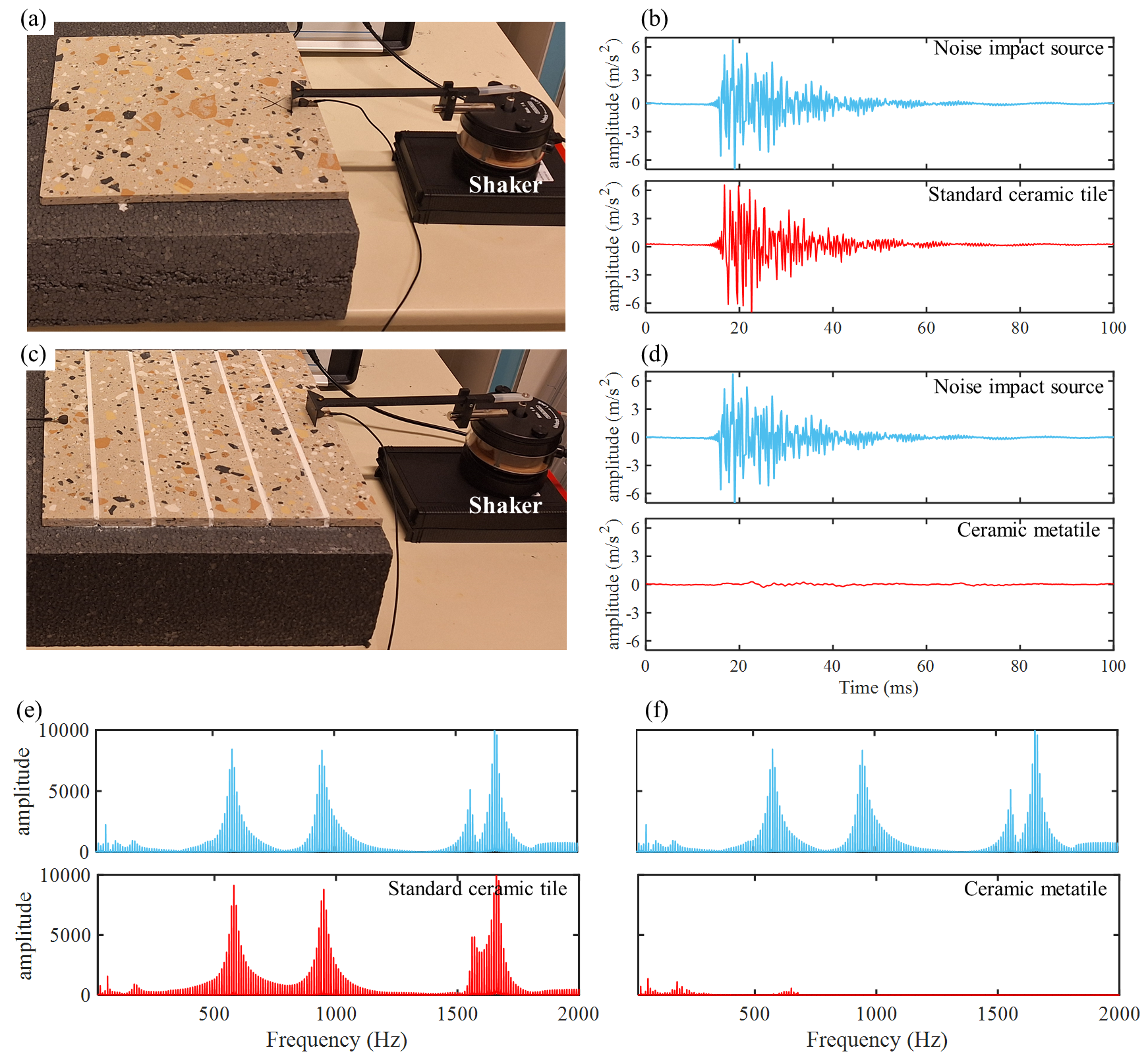}
    \caption{Impact noise application using a shaker along the [100] direction. (a) Standard ceramic tile subjected to an impact generated by a shaker. (b) Time response of the standard ceramic tile: the blue curve represents the measured impact noise source, while the red curve corresponds to the transmitted noise on the opposite side of the tile. (c) Ceramic metatile subjected to an impact generated by a shaker. (d) Time response of the ceramic metatile: the blue curve represents the measured impact noise source, while the red curve corresponds to the transmitted noise on the opposite side of the metatile. (e) and (f) FFT analysis of the source and transmitted vibrations for the standard ceramic tile and the ceramic metatile, respectively.}
    \label{figure 8}
\end{figure*}

After numerically and experimentally demonstrating the existence of stop bandgaps in the ceramic metatile, this section focuses on studying impact noise reduction in both the standard ceramic tile and the metatile using a real impact noise source. The setup, illustrated in \autoref{figure 8}(a) and (c), involves a shaker connected to a hammer-like object to simulate the impact noise generated by footsteps on tiled floors.
We excite the shaker by a theoretical signal, which is a Gaussian-like signal with an interval of $100$ ms and we employ the same laboratory setup as depicted in \autoref{figure 2}.

A transducer is placed to measure the impact noise source, while another transducer is placed on the opposite side of the tiles in order to measure their transmission noise amplitude. The results of this study are shown in \autoref{figure 8} (b) and (d), which show the time-dependent acceleration amplitude for both the standard ceramic tile and the metatile. The impact noise source is represented by the blue curve, while the transmitted noise is shown in red. Standard ceramic tile exhibits high transmitted noise, characterized by a large amplitude over time. In contrast, the metatile shows a significantly lower amplitude, indicating its effectiveness in attenuating noise propagation.

To further analyze this behavior, the Fast Fourier Transform (FFT) of both responses is calculated and presented in \autoref{figure 8}(e) and (f). The standard ceramic tile allows noise transmission across the frequency spectrum from $20$ Hz to $2000$ Hz. However, the metatile demonstrates a significant reduction in noise transmission at low frequencies, and beyond $500$ Hz, no impact noise transmission is observed. These results highlight the strong noise control and attenuation capabilities of the metatile, making it a promising solution for tiled flooring applications.

The comparison between the standard ceramic tile and our newly proposed design, referred to as the ceramic metatile, demonstrates that the inclusion of silicon joints significantly enhances the elastodynamic behavior of the tile. This enhancement is evident in the emergence of bandgaps and the improved attenuation of impact noise on tiled floors. Notably, the materials used in the ceramic metatile are the same as those in standard ceramic tiles, and the use of silicon adhesive is already common in real-world applications for joining tiles. Our study shows that, by combining these existing materials with structural engineering principles, it is possible to substantially improve impact noise and vibration comfort in tiled rooms.

In terms of cost and material effectiveness, the proposed ceramic metatile remains comparable to standard ceramic tiles. The only modification involves using sub-tiles with dimensions of $5 \times 30 cm^2$. These can be produced either by designing a new mold or by cutting existing standard tiles using a ceramic cutter. The silicon joints, already used in tile installations, can be repurposed in the same manner for constructing the metatile. Regarding long-term durability, the metatile is expected to perform similarly to standard vitrified ceramics, which are well-known for their long-lasting performance. As for installation, the process is straightforward and compatible with conventional methods. One approach involves placing the sub-tiles individually and applying the silicon joints between them to form the metatile directly on-site. Alternatively, preassembled ceramic metatiles could be manufactured and installed using magnetic backing sheets, offering easy installation and replacement. In summary, our design not only exhibits strong attenuation of elastic waves but also holds promise for further optimization to target lower frequencies, potentially below $100$ Hz, while maintaining compatibility with current building practices and materials.

{Additionally, further design strategies based on the use of silicon joints or other soft materials can enhance the insulation performance of the tiled floor. In this study, we presented a unidirectional arrangement of the sub-tiles and silicon joints. However, an alternative configuration is to arrange the sub-tiles and silicon joints along both the $x$- and $y$-directions (see Appendix \ref{apC}), which could further improve attenuation in both directions of the tiled floor.}

\section{Conclusion}
{To conclude, we presented novel designs of ceramic metatiles that combine ceramic tiles with silicon joints. To characterize their elastic insulation behavior, we performed numerical optimizations and calculated their transmission and dispersion curves in order to optimize the dynamic response of the designs. Subsequently, we validated our numerical approach through experimental testing. For that, we fabricated homemade ceramic metatiles and employed a laboratory setup using Fast Fourier Transformation to measure their transmission responses. Furthermore, we highlighted the standard acoustic properties of standard ceramic tiles, commonly used in tiled floors, as a reference to contrast with the outcomes obtained from our ceramic metatiles. As a founding, we found broadband bandgaps at low-frequency ranges for both flexural and longitudinal waves, in contrast to standard tiles where such features are less pronounced for elastic waves. Our findings showed that for {flexural} waves, the ceramic metatile exhibits a significant reduction in transmission within the $500$–$1900$ Hz range along the $[100]$ direction and within the $500$–$1400$ Hz range along the $[110]$ direction. For longitudinal waves, the ceramic metatile displays a broad bandgap extending from $400$ Hz to $1950$ Hz along both the $[100]$ and $[110]$ directions. In addition, we observed a shift of the bandgaps toward lower frequencies as the width of the subtiles and silicon joints increases, or as the Young’s modulus of silicon decreases. This indicates that the stop bands can be tailored through geometric parameters and material properties to effectively target low-frequency ranges, such as those associated with footstep and indoor noise propagating through elastic waves in tiled floors. Finally, the homemade ceramic metatile samples confirmed the numerical results, the experimental and numerical data are in good agreement, both showing strong elastic stop bands for elastic impact noise. This validates the efficiency and reliability of our design in attenuating elastic wave propagation through tiled floors. Overall, we demonstrated the potential of our novel ceramic metatile design for practical applications in architectural acoustics, particularly in controlling unwanted vibrations and noise in building indoors.}
\section*{Acknowledgements}
This research was funded by IRIS, BTU Trust, the French ANR (contract "ANR-21-CE33-0015") and the European Union under Marie Skłodowska-Curie Actions Postdoctoral Fellowships (No. 101149710). The authors acknowledge the support of French-Swiss SMYLE Network. We thank Julien Joly (UFR ST, Dep Elec, UMLP) for 3D printed parts.

\appendix
\section{{Calculation of the absolute error between experimental and numerical flexural wave transmissions along the $[100]$ direction.}}\label{apA}
\begin{figure}[h]
    \centering
    \includegraphics[width=0.5\linewidth]{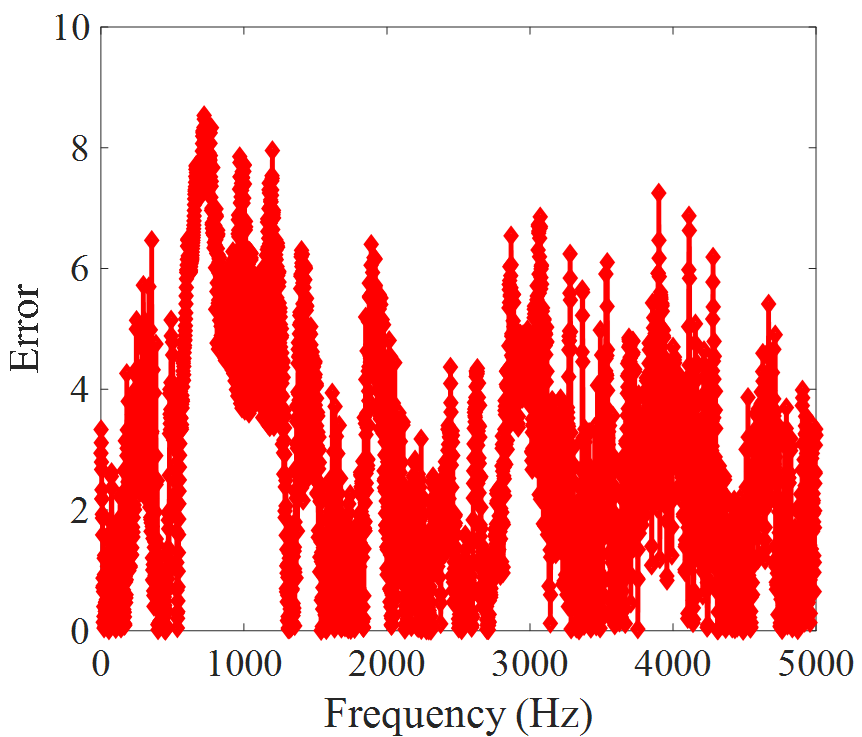}
    \caption{{Absolute error versus frequency for flexural wave transmissions along the $[100]$ direction.}}
    \label{error}
\end{figure}
\section{{Calculation of damping effects on elastic wave transmission through silicon joints.}}\label{apB}
\begin{figure}[h]
    \centering
    \includegraphics[width=0.5\linewidth]{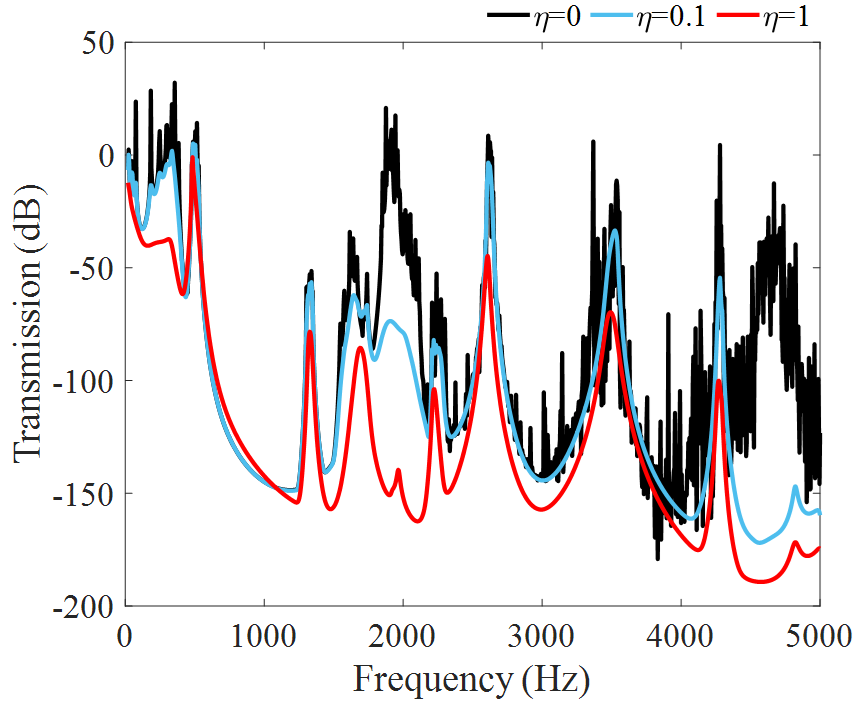}
    \caption{{Flexural wave transmission along the [100] direction for three loss factor cases: black for $\eta=0$, blue for $\eta=0.1$ and red for $\eta=1$.}}
    \label{damping}
\end{figure}

\section{{Designs of ceramic metatiles with omnidirectional broadband band gaps.}}\label{apC}
\begin{figure}[h]
    \centering
    \includegraphics[width=0.5\linewidth]{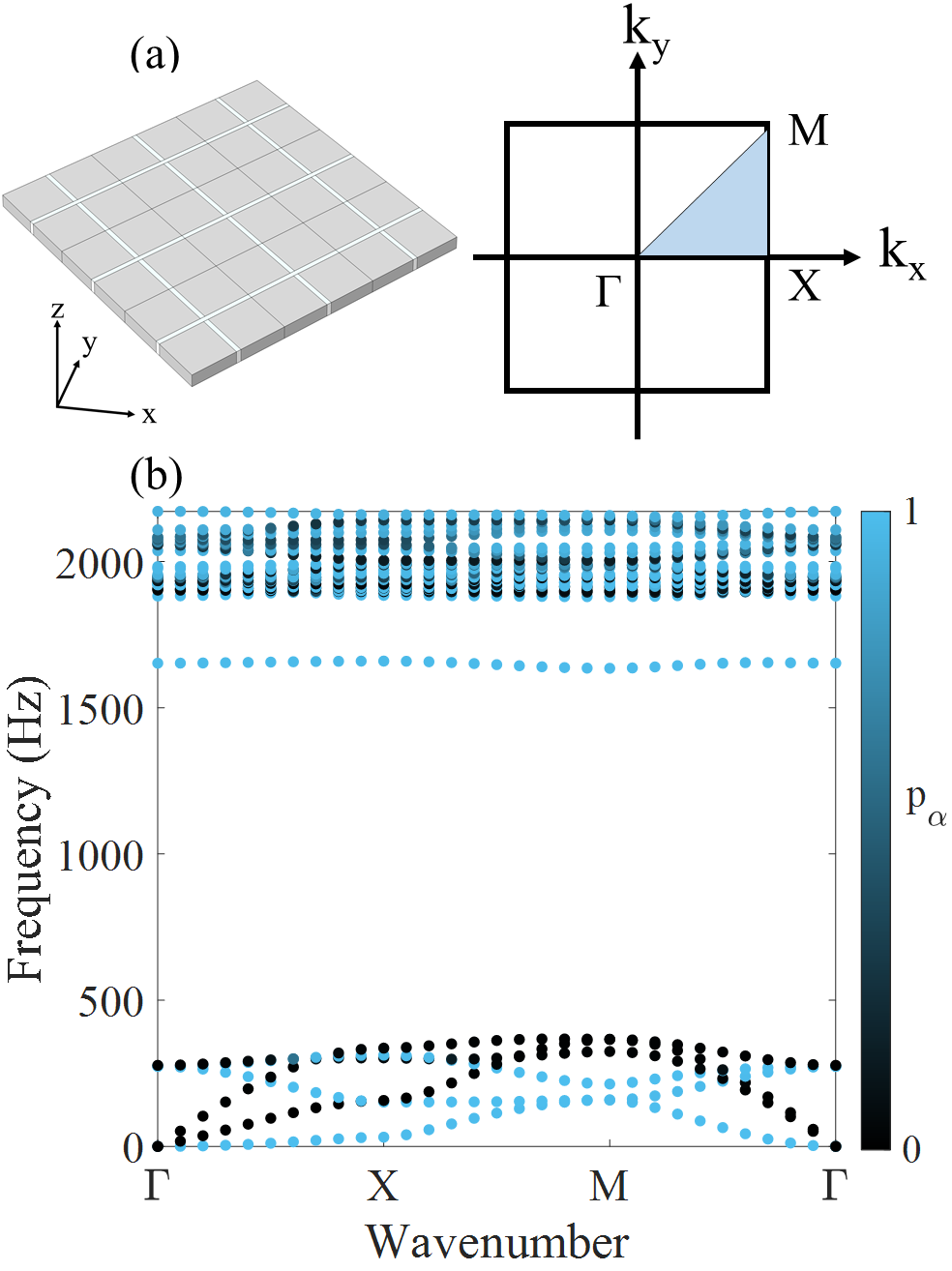}
    \caption{{(a) Rectangular sub-tiles with silicon joints arranged along both the $x$- and $y$-directions, and the corresponding first irreducible Brillouin zone ($\Gamma XM\Gamma$), with points defined as $\Gamma = (0, 0, 0)$, $X = (\pi/d, 0, 0)$, and $M = (\pi/d, \pi/d, 0)$.. b) Calculated phononic dispersion in the first irreducible Brillouin zone ($\Gamma XM\Gamma$).}}
    \label{configurations}
\end{figure}

\bibliographystyle{MSP}
\bibliography{Mybib}

\end{document}